\documentclass[letterpaper, 10pt, conference]{ieeeconf}      

\IEEEoverridecommandlockouts                            
\usepackage[usenames,dvipsnames]{xcolor}
\usepackage{colortbl}
\usepackage{graphicx} 
\usepackage{mathptmx} 
\usepackage{times} 
\usepackage{amsmath} 
\usepackage{amssymb}  
\usepackage{pstricks,pst-plot,psfrag}
\usepackage{units}
\usepackage{multirow}
\usepackage{hyperref}
\usepackage{booktabs}
\usepackage{balance}
\usepackage[acronym]{glossaries}

\usepackage{lettrine}
\usepackage{import}
\usepackage{tikz}
\usepackage{lipsum}  
\usetikzlibrary{positioning}
\usetikzlibrary{calc}

\definecolor{lightblue}{rgb}{0.60784,0.76078,0.90196}
\definecolor{darkblue}{rgb}{0.26667,0.44706,0.76863}
\definecolor{lightgreen}{rgb}{0.66275,0.81569,0.55686}
\definecolor{darkgreen}{rgb}{0.43922,0.67843,0.27843}
\definecolor{orange}{rgb}{0.92941,0.49020,0.19216}
\definecolor{yellow}{rgb}{1.00000,0.75294,0.00000}
\definecolor{grey}{rgb}{0.64706,0.64706,0.64706}
\definecolor{purple}{rgb}{0.51373,0.23529,0.04706}
\newacronym{abk:amod}{AMoD}{Autonomous Mobility-on-Demand}
\newacronym{abk:iamod}{I-AMoD}{intermodal \gls{abk:amod}}
\newacronym{abk:bpr}{BPR}{the Bureau of Public Roads}
\newacronym{abk:ca}{CA}{congestion-aware}
\newacronym{abk:cara}{CARS}{congestion-aware routing scheme}
\newacronym{abk:cepamods}{CEPAMoDS}{Convolutional Energy Predicting AMoD Scheduler}
\newacronym{abk:cs}{CS}{charging station}
\newacronym{abk:ffcs}{FFCS}{free floating car sharing systems}
\newacronym{abk:ghg}{GHG}{greenhouse gas}
\newacronym{abk:kpi}{KPIs}{Key Performance Indicators}
\newacronym{abk:mcfp}{MCFP}{multi-commodity flow problem}
\newacronym{abk:milp}{MILP}{mixed-integer linear program}
\newacronym{abk:spp}{SPP}{shortest path problem}
\newacronym{abk:kdspp}{k-dSPP}{k-disjoint \gls{abk:spp}}
\newacronym{abk:soc}{SoC}{state of charge}
\newacronym{abk:vrp}{VRP}{vehicle routing problem}
\newacronym{abk:v2g}{V2G}{vehicle-to-grid}
\newcommand{\flexbrac}[1]{\if\relax\detokenize{#1}\relax \else (#1) \fi}
\newcommand{\flexcomma}[1]{\if\relax\detokenize{#1}\relax \else ,#1 \fi}

\newcommand{\sR}{\mathbb{R}}

\newcommand{\ddt}{\frac{\textnormal{d}}{\textnormal{d}t}}

\usepackage{mathtools}
\usepackage{dsfont}
\usepackage{booktabs}
\usepackage{bbm}
\usepackage{diagbox}
\usepackage{cuted}
\usepackage{soul}
\usepackage{comment}
\let\proof\relax
\let\endproof\relax

\usepackage{amsthm}
\usepackage[font=scriptsize]{caption}
\usepackage[aboveskip=10pt,font=scriptsize, belowskip=0pt]{subcaption} 
\usepackage{array}

\newtheorem{theorem}{Theorem}[section]

\makeatletter
\newcommand{\pushright}[1]{\ifmeasuring@#1\else\omit\hfill$\displaystyle#1$\fi\ignorespaces}
\newcommand{\pushleft}[1]{\ifmeasuring@#1\else\omit$\displaystyle#1$\hfill\fi\ignorespaces}
\newcolumntype{x}[1]{>{\centering\arraybackslash\hspace{0pt}}p{#1}}
\makeatother
\newif\ifmargincomments 
\margincommentstrue

\newif\ifextendedversion 
\extendedversionfalse

\ifmargincomments
\else

\fi

\newif\ifsuggestions
\suggestionstrue

\ifsuggestions
\newcommand{\stst}[2]{{\color{orange}\st{#1}}{\color{orange}#2}}

\else
\newcommand{\stst}[2]{}
\fi

\title{\LARGE \bf
	Optimizing Vaccine Allocation Strategies in Pandemic Outbreaks:\\ An Optimal Control Approach
}

\author{Sander Tonkens$^{1,*}$, Paul de Klaver${^2}$, and Mauro Salazar$^{3}$
	\thanks{$^1$Sander Tonkens is with the Department of Mechanical and Aerospace Engineering, University of California, San Diego, United States, e-mail:
		{\tt\small stonkens@ucsd.edu}.}
	\thanks{$^2$Paul de Klaver is with the M\'axima Medisch Centrum, Eindhoven, The Netherlands, email:
		{\tt\small p.deklaver@mmc.nl}.}
	\thanks{$^3$Mauro Salazar is with the Control Systems Technology section, Department of Mechanical Engineering, Eindhoven University of Technology, The Netherlands, e-mail:
		{\tt\small m.r.u.salazar@tue.nl}.}
	\thanks{$^*$Work done while working at Eindhoven University of Technology.}
}

\graphicspath{{./img/}}
\begin{document}
	\maketitle

	\begin{abstract}
		Since early 2020, the world has been dealing with a raging pandemic outbreak: COVID-19.
		A year later, vaccines have become accessible, but in limited quantities,
		so that governments needed to devise a strategy to decide which part of the population to prioritize when assigning the available doses, and how to manage the interval between doses for multi-dose vaccines.
		In this paper, we present an optimization framework to address the dynamic double-dose vaccine allocation problem whereby the available vaccine doses must be administered to different age-groups to minimize specific societal objectives.
		In particular, we first identify an age-dependent Susceptible-Exposed-Infected-Recovered (SEIR) epidemic model including an extension capturing partially and fully vaccinated people, whereby we account for age-dependent immunity and infectiousness levels together with disease severity.
		Second, we leverage our model to frame the dynamic age-dependent vaccine allocation problem for different societal objectives, such as the minimization of infections or fatalities, and solve it with nonlinear programming techniques.
		Finally, we carry out a numerical case study with real-world data from The Netherlands.
		Our results show how different societal objectives can significantly alter the optimal vaccine allocation strategy.
		For instance, we find that minimizing the overall number of infections results in delaying second doses, whilst to minimize fatalities it is important to fully vaccinate the elderly first.
	\end{abstract}

	\section{Introduction}\label{sec:introduction}
Since the start of 2020, the world has been facing a pandemic with devastating effects on society: As of September 2021, more than 225 million infections and 4.6 million deaths have been directly attributed to the severe acute respiratory syndrome coronavirus 2 (SARS-CoV-2)~\cite{DongDuEtAl2020}.
To fight this pandemic, a large number of non-pharmaceutical interventions (NPIs), e.g., stay-at-home orders and public mask-wearing have been put in place to limit the spread of the virus. Additionally, pharmaceutical interventions, such as immunosuppressant drugs, have led to a reduced fatality-per-infection rate compared to Spring 2020. While these methods have shown to be essential in somewhat containing the spread of the virus and limiting its impacts, vaccines seem to be the only long-term solution~\cite{MooreHillEtAl2021b}.

\begin{figure}[t]
	\includegraphics[width = \columnwidth]{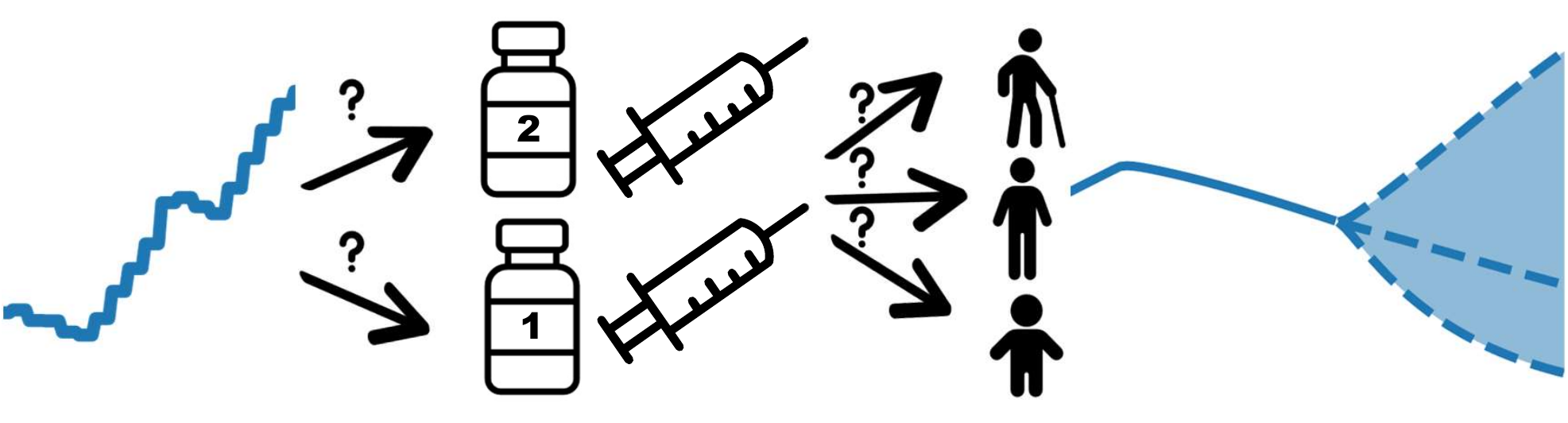}
	\caption{In this work we devise a framework to solve the dynamic vaccine allocation problem. Importantly, this framework allows prioritizing certain age groups (based on the desired objective) and between partially and fully vaccinating individuals. The limited initial availability of vaccines adds an additional dimension to the problem. Different strategies can have big impacts on the outcome (depicted on the right graph).}
	\label{fig:first_page}
	\vspace{-2em}
\end{figure}

As vaccines are becoming available, but in limited quantities, governments need to decide which strategies to adopt in terms of how to distribute the doses to limit the outbreak and minimize societal damage.
Crucially, whilst already providing some level of immunity after one inoculation, most of the approved vaccines must be administered multiple times.
Therefore it is important to understand which population groups should be vaccinated first and whether the subsequent doses should be reserved and administered as soon as possible (e.g., after three weeks) or given as a first dose to another individual.
Specifically, given a pre-defined societal cost such as number of total infections or number of casualties, when and to what extent should an age-group be vaccinated in order to minimize such a cost?
Assuming the societal cost to be given (e.g., by a transdisciplinary panel of experts), in this paper we devise a modeling and optimization framework to solve the optimal dynamic vaccine allocation problem with different age-groups that is based on a multi-state Susceptible-Exposed-Infected-Recovered (SEIR) model accounting for vaccinated people.

\emph{Related work:}\label{par:related_work}
This work is related to two research areas: the modeling of infectious diseases and vaccine allocation, with a special focus on COVID-19. We proceed by briefly discussing related work on modeling COVID-19. Following this, we discuss the vaccine allocation problem, where we focus on age-based prioritization, as susceptibility and disease severity have been shown to be very strongly correlated to age for COVID-19~\cite{DaviesKlepacEtAl2020,MuellerMcNamaraEtAl2020,LevinHanageEtAl2020}.

To account for factors that are specific to COVID-19, such as pre- and asymptomatic spread and government interventions, many extensions to the classical SEIR compartmental model have been proposed. Crucially, very limited testing capacity at the start of the pandemic required capturing both diagnosed and not diagnosed individuals~\cite{GiordanoBlanchiniEtAl2020} and fine grained models to predict superspreader events~\cite{ChangPiersonEtAl2021}. Early on in the pandemic, many government-instituted taskforces devised stochastic SEIR-extended models (e.g.,~\cite{AbramsWambuaEtAl2021}) to persuade their respective governments to impose restrictive social-distancing measures. These stochastic compartmental models can capture infection dynamics at a very localized level, but are computationally intractable when implemented in an optimization framework.

The majority of pre-COVID-19 studies on vaccine allocation focus on distributing a fixed number of vaccines over a population to suppress an infectious disease, leveraging either static optimization or simulations of pre-defined distribution strategies. We refer the reader to~\cite{DuijzerJaarsveldEtAl2018b} for an overview. Framing the problem as a static optimization problem allows to obtain theoretical results~\cite{DuijzerJaarsveldEtAl2018} which demonstrate the advantage of prioritizing a group-by-group approach for non-mixing populations over proportional allocation.
Nonetheless, due to the high infectiousness, the devastating effects of COVID-19 on society, and the limited availability of vaccines, it is essential to consider dynamic optimization to reason about vaccine allocation over time and under (severe) and varying supply constraints. Approaches which consider vaccines to have been administered \emph{a priori} (e.g.,~\cite{MatrajtEatonEtAl2021b}) fail to capture how vaccinated people impact the disease dynamics during the vaccination campaign.
Within dynamic vaccine allocation we equally distinguish between simulation-based and dynamic optimization approaches. Simulation-based approaches evaluate the effect of different pre-defined prioritization strategies. For example, the authors of~\cite{BubarReinholtEtAl2021} simulate five age-based prioritization strategies for different levels of total vaccine supply (demonstrating the need to prioritize vaccinating the elderly to minimize mortality). However, simulation-based approaches are limited to pre-defined strategies, and hence do not enable to fully optimize the allocation strategies.
Conversely, existing dynamic optimization-based approaches rely on linearized models and convex optimization~\cite{SmithBullo2021}, or nonlinear optimization techniques, such as coordinate descent~\cite{BertsimasIvanhoeEtAl2020} and genetic algorithms~\cite{BucknerChowellEtAl2021} to find a local optimum. Yet they are limited to single-dose vaccinations that are immediately effective, whilst multi-dose vaccines have not been studied in depth yet.
In this context, the authors~\cite{MatrajtEatonEtAl2021} show the importance that delaying second doses can have, but only consider pre-defined strategies that always prioritize old-to-young when new vaccines become available.
Hence, it is important to capture both the requirement of having multiple vaccine doses to be fully vaccinated and not limit the optimization space by enforcing pre-defined prioritization schemes.

In summary, prior work is limited to either evaluating pre-defined strategies, failing to capture the effect that delaying second doses can have, or considering objectives beyond minimizing mortality, years of life lost (YLL) and infections, such as hospitalization levels and intensive care (IC) pressure. Additionally, many of these works use models that fail to capture the actual disease dynamics, by considering models parameterized using early pandemic data and do not consider the dynamic evolution of the pandemic over time.

\emph{Statement of Contributions:}\label{par:statement_of_contributions}
In this work, we present an optimization framework for dynamic age-based vaccine allocation. Specifically, we devise an age-stratified extended-SEIR model including vaccinated people (SEIR-V\textsubscript{2}) that explicitly accounts for variable dosing intervals for multi-dose vaccinations, vaccine hesitancy, and limited vaccine availability. The presented framework enables optimizing for (read: minimizing) a wide variety of objectives, including cumulative outcomes and daily IC and hospital occupancy. We demonstrate the utility of the framework by \emph{i)} fitting the model to real-world data describing the evolution of the pandemic in The Netherlands over a 1 year period, and \emph{ii)} with the fitted model, we compare outcomes based on different objectives. We focus on two case studies: optimizing for minimizing infections and minimizing fatalities, thereby comparing the different strategies obtained.

\emph{Disclaimer:}\label{par:disclaimer}
The objective of this paper is to provide insights to policymakers and practitioners through a numerical framework by framing the optimal vaccine allocation problem and solving it for a range of objectives.
Our work does not offer a recommendation on how to prioritize vaccine allocation nor provide any political statement.
As in any research paper, our results should be interpreted accounting for the underlying assumptions and the available model accuracy.

\emph{Organization:}\label{par:organization}
The remainder of this paper is structured as follows. Section~\ref{sec:methodology} introduces the standard SEIR epidemiological model and extends it to an age-stratified SEIR-V\textsubscript{2} model, before leveraging this model to frame the optimal dynamic vaccine allocation problem.
We identify the model's parameters for The Netherlands in Section~\ref{sec:fitting_results} and use the identified model to solve the optimal allocation problem for different objectives in Section~\ref{sec:optimization_results}. We draw the conclusions and discuss future work in Section~\ref{sec:conclusion}.

	\section{Methodology}\label{sec:methodology}
In this section, we first devise an age-stratified SEIR model capturing two-dose vaccinations in Section~\ref{subsec:compartmental_model}, and then leverage it to frame the optimal dynamic vaccine allocation problem in Section~\ref{subsec:ocp}. We conclude with a discussion of our approach in Section~\ref{subsec:model_discussion}.
\subsection{Compartmental Models}\label{subsec:compartmental_model}
Considering a population whereby $s(t)$ represents the share of susceptible individuals, $e(t)$ the share of the population that is exposed, but not yet infectious, $i(t)$ the share that is currently infected, and $r(t)$ the share that has recovered (including severe outcomes), the SEIR model~\cite{BrauerDriesscheEtAl2008} is as follows:
\begin{subequations}\label{eq:SIR_model}
	\begin{align}
		\ddt s(t) &= -\beta s(t) i(t) \\
		\ddt e(t) &= \beta s(t) i(t) - \gamma_\mathrm{e} e(t) \\
		\ddt i(t) &= \gamma_\mathrm{e} e(t) - \gamma_\mathrm{i} i(t) \\
		\ddt r(t) &= \gamma_\mathrm{i} i(t),
	\end{align}
\end{subequations}
with $\beta$ the infection rate, $1/\gamma_\mathrm{e}$ is the latent period, $\gamma_\mathrm{i}$ the recovery rate ($1/\gamma_\mathrm{i}$ is the average infectious period). As $s(t)$, $i(t)$, $e(t)$, and $r(t)$ represent shares of the population, we have $s(t) + e(t) + i(t) + r(t) = 1$, with $s(t)$, $e(t)$, $i(t)$, \mbox{$r(t) \in [0, 1]$}. As a side note, the widely reported base reproduction rate $R_0 = \frac{\beta}{\gamma_\mathrm{i}}$ and the actual reproduction rate $R(t) = R_0 s(t)$ are calculated based on the parameters of the SEIR model. We would like to highlight that the SEIR model, although it only captures the high-level disease dynamics, involves a cross-term that can result in a highly nonlinear behavior.
Finally, we do not consider vital dynamics (birth and death) as they are much slower than the dynamics of the epidemic.

Whilst this model is useful to gain an understanding of the spread of infectious diseases and for performing theoretical analyses, e.g.,~\cite{DuijzerJaarsveldEtAl2018}, it is typically extended for use in the forecasting and simulation of an infectious disease. In our work, we address some key shortcomings of the SEIR model by proposing the following extensions:
\begin{enumerate}
	\item We consider an age-stratified model to capture both non-homogeneous spread (e.g., a 20-year-old is more likely to interact with a 25-year-old than an 80-year-old) and to allow allocating the vaccines at unequal rates to different age groups. Hereby, the contact matrix $c$ indicates the level of direct interactions within and between groups to drive the transmission dynamics in the model. 
	\item We consider hospitalizations and fatalities as a subset of the recovered individuals to take severe outcomes and hospital occupancy into account.
	\item We scale the transmission rate by a time-varying factor $\theta(t) \in [0, 1]$ to reflect the effect of NPIs and other transmission-limiting effects such as human behavior (e.g., pandemic fatigue) and seasonality. Specifically, we consider a piecewise constant $\theta(t)$. Notably, to avoid overfitting, e.g., by fitting to a new $\theta(t)$ every day, we split the domains based on changes in policy measures.
	\item We incorporate single and fully vaccinated equivalents of the susceptible, exposed, infected, and recovered states. Our approach to modeling vaccinated individuals, i.e., reducing the probability that susceptible individuals will be infected, is commonly referred to as a \emph{leaky} vaccine model.
\end{enumerate}
Combined, the aforementioned extensions allow us to define a population where $s_i^k(t)$, $e_i^k(t)$, $i_i^k(t)$, and $r_i^k(t)$ define the share of susceptible, exposed, infected, and recovered individuals for each vaccination status $k=\{\square, \mathrm{v_1}, \mathrm{v_2}\}$, i.e., not vaccinated, single dose, fully vaccinated, and each age group~$i$. Furthermore, we define the inputs $u_i^{\mathrm{v_1}}(t)$ and $u_i^{\mathrm{v_2}}(t)$ as the share of individuals (proportional to the entire population) in group $i$ that get their first or second dose at time $t$. The SEIR-V\textsubscript{2} model is as follows (where we drop time-dependency for the sake of readability):
\begingroup
	\allowdisplaybreaks
	\begin{subequations}\label{eq:dynamics}
		\begin{align}
			\ddt s_i =& -\beta_i s_i - u_i^{\mathrm{v_1}}\frac{s_i}{s_i + r_i} \\
			\ddt e_i^k =& (1-\eta_{\mathrm{s},i}^{k})\beta_i s_i^k - \gamma_\mathrm{e} e_i^k \\
			\ddt i_i^k =& \gamma_\mathrm{e} e_i^k - \gamma_\mathrm{i} i_i^k \\
			\ddt r_i =& \gamma_\mathrm{i} i_i^k - u_i^{\mathrm{v_1}}\frac{r_i}{s_i + r_i} \\
			\ddt s_i^{\mathrm{v_1}} =& -(1-\eta_{\mathrm{s},i}^{\mathrm{v_1}})\beta_i s_i^{\mathrm{v_1}} + u_i^{\mathrm{v_1}}\frac{s_i}{s_i + r_i} \nonumber\\* &- u_i^{\mathrm{v_2}} \frac{s_i^{\mathrm{v_1}}}{s_i^{\mathrm{v_1}} + r_i^{\mathrm{v_1}}}\\
			\ddt s_i^{\mathrm{v_2}} =& -(1-\eta_{\mathrm{s},i}^{\mathrm{v_2}})\beta_i s_i^{\mathrm{v_2}} + u_i^{\mathrm{v_2}} \frac{s_i^{\mathrm{v_1}}}{s_i^{\mathrm{v_1}} + r_i^{\mathrm{v_1}}} \\
			\ddt r_i^{\mathrm{v_1}} =& \gamma_\mathrm{i} i_i^{\mathrm{v_1}} + u_i^{\mathrm{v_1}}\frac{r_i}{s_i + r_i} - u_i^{\mathrm{v_2}} \frac{r_i^{\mathrm{v_1}}}{s_i^{\mathrm{v_1}} + r_i^{\mathrm{v_1}}} \\
			\ddt r_i^{\mathrm{v_2}} =& \gamma_\mathrm{i} i_i^{\mathrm{v_2}} + u_i^{\mathrm{v_2}} \frac{r_i^{\mathrm{v_1}}}{s_i^{\mathrm{v_1}} + r_i^{\mathrm{v_1}}},
		\end{align}
	\end{subequations}
\endgroup
for all groups $i$ and $k \in \{\square, \mathrm{v_1}, \mathrm{v_2}\}$. The age-dependent time-varying infection rate is $\beta_i(t) = \theta(t) \zeta_i \sum\limits_j \left(c_{i,j}\sum\limits_k\left(1 - \eta_{\text{inf},j}^k\right) i_j^k\right)$, where $\zeta_i$ is the probability of a successful transmission given contact with an infectious individual and the contact matrix $c_{i,j}$ captures the number of contacts of an individual of group $i$ with one of group~$j$. We account for the vaccine's effectiveness in reducing susceptibility $\eta_{\mathrm{s},i}$ and infectiousness $\eta_{\text{inf}, i}$. The combined state is given by $x \in \mathbb{R}^{4 \cdot 3 \cdot N_\mathrm{groups}}$, with $N_\mathrm{groups}$ the number of age-groups. As for~\eqref{eq:SIR_model}, we have $\sum\limits_i x_i(t) = 1$ and $x_i(t) \in [0, 1]$ for all $t$.

\subsection{Optimal Control Problem}\label{subsec:ocp}
We can now embed the aforementioned model---Eq.~\eqref{eq:dynamics}---in a constrained optimization problem. We minimize a total cost $J$ that can represent different objectives, such as total number of fatalities or maximum hospital occupancy. We proceed by discussing the constraints we impose, followed by the definition of the full optimization problem.

First and foremost, we consider a time-varying bound on the overall vaccine supply and enforce positivity constraints on the inputs:
\begin{align}\label{eq:umax}
	\begin{split}
		\sum_i u_i^{\mathrm{v_1}}(t) + u_i^{\mathrm{v_2}}(t) &\leq u_{\text{max}}(t) \\ 
		u_i^{\mathrm{v_1}}(t), u_i^{\mathrm{v_2}}(t) &\geq 0  \ \forall \ i, t.
	\end{split}
\end{align}
Next, we introduce a bound on the minimum and maximum delay between first and second doses. A minimum delay $\underline{{D}}$ is often strictly recommended by the vaccine manufacturer, whilst exceeding a maximum delay $\overline{{D}}$ between the first and second dose might adversely impact immunity due to a loss of immunological memory:
\begin{align}\label{eq:delays}
	\begin{split}
		u_i^{\mathrm{v_2}}(t) &\leq V^{\mathrm{v_1}}_i(t-\underline{{D}}) + 		V^{\mathrm{v_2}}_i(t-\underline{{D}}) - V^{\mathrm{v_2}}_i(t)\\
		u_i^{\mathrm{v_2}}(t) &\geq V^{\mathrm{v_1}}_i(t-\overline{{D}}) + 	V^{\mathrm{v_2}}_i(t-\overline{{D}}) - V^{\mathrm{v_2}}_i(t),
	\end{split}
\end{align}
with $V^k_i = s^k_i + e^k_i + i^k_i + r^k_i$.

Lastly, we consider that a share of people $\xi_i \in [0,1]$ does not get the vaccine  for each age group $i$, both voluntarily and/or due to medical restrictions:
\begin{align}\label{eq:hesitancy}
	\sum_t u_i^{\mathrm{v_1}}(t) \leq (1- \xi_i) \frac{{N}_i}{{N}_{\text{tot}}} \ \forall \ i, t,
\end{align}
with ${N}_i$ denoting the population of group $i$ and ${N}_\text{tot} = \sum\limits_i {N}_i$ the total population.
With this model in place, we have all the ingredients to present the optimal control problem, where $u(t)=[{u^{\mathrm{v_1}}}(t), {u^{\mathrm{v_2}}}(t)]^\top \in \sR^{2 N_\mathrm{groups}}$ are the input variables:
\begin{align}
	\begin{split}\label{eq:ocp}
		\underset{u}{\text{minimize}} \quad & J\\
		\text{subject to} \; \; \; \:  & \eqref{eq:dynamics}, \eqref{eq:umax}, 	\eqref{eq:delays}, \eqref{eq:hesitancy}.
	\end{split}
\end{align}

\subsection{Discussion}\label{subsec:model_discussion}
A few comments are in order.
First, we do not consider the waning efficacy of the vaccine, nor the possibility of losing immunity from prior infection over time, and we assume immediate immunity after each dose. Yet our framework can be readily extended to account for such phenomena as soon as sufficient data are available.
Second, we do not explicitly model different variants of the virus. However, these are indirectly reflected through the value of the contact reduction parameter $\theta(t)$.
Third, we do not explicitly capture risk groups, e.g., immunodepressed people or essential healthcare workers, and assume they represent a non-significant share of each age category. However, whilst the former group can be directly considered as part of the elderly population, both groups can be readily included via additional variables.
Lastly, the solution found with our framework is not guaranteed to be globally optimal due to the inherently non-convex nature of our model---see Eq.~\eqref{eq:dynamics}. This issue is present in even the simplest compartmental model, as the rate of newly infected people is a bilinear function of the currently susceptible and infected populations---see Eq.~\eqref{eq:SIR_model}---and cannot be completely overcome. Against this backdrop, we can use different warm-starting points to empirically assess the quality of the obtained local optima~\cite[Chpt.~1]{Bertsekas2016}.

	\section{Fitting to Real-world Data}\label{sec:fitting_results}
In this section, we first describe the data that we use to identify our model parameters. We then elaborate on which parameters, see Eq.~\eqref{eq:dynamics}, are considered fixed, whilst detailing the bounds for the parameters to be fitted.

As mentioned in the related works in Section~\ref{sec:introduction}, stochastic compartmental models can capture the disease dynamics better, both in accounting for uncertainty and individual-level spread. However, as we argued before, these models are computationally intractable for use in an optimization framework. Instead, we use the deterministic model, Eq.~\eqref{eq:dynamics}, to fit to a stochastic model developed by the Dutch National Institute for Public Health and the Environment (RIVM)~\cite{RIVM:2021c}. Specifically, we fit the number of infectious people at a given point in time in our model to the estimated prevalence from~\cite{RIVM:2021c}. By doing so, we avoid the need to consider a more complicated model that captures non-detected infections and distinguishes asymptomatic and symptomatic individuals. 
While being more accurate, the RIVM model, unlike Eq.~\eqref{eq:dynamics}, only captures the overall prevalence. To fit the respective age groups to the prevalence data, due to the age-segmented prevalence data not being open-sourced, we assume that the share of positive tests per age group is proportional to the share of infectious people per age group. 

In addition to prevalence data, we capture the impact of vaccinations to the disease dynamics in The Netherlands by considering the number of people vaccinated by the Dutch government per age group~\cite{RIVM:2021b} as an input to the model. 

To fit our model, Eq.~\eqref{eq:dynamics}, to the daily reported vaccinations, case data, and estimated prevalence we discretize Eq.~\eqref{eq:dynamics} using Euler Forward with a sampling period of 1 day. We fit the model over the period of one year, from July 2020 to June 2021. Given the high uncertainty of the disease dynamics and the low availability of tests during the first wave in Spring 2020, we initialize our model after the end of the first wave, and set the initial value of recovered individuals at the start to be $5\%$, in line with estimates from the RIVM~\cite{RIVM:2021c}.
Moreover, we fix the initial value of the number of infected individuals to be equal to the estimated prevalence from~\cite{RIVM:2021}, and consider the number of exposed individuals to be half of that number. The susceptible share sums the model to 1, and all vaccinated states have an initial value of 0.

\begin{figure}[!t]
	\centering
	\setlength\tabcolsep{0.7pt}
	\renewcommand{\arraystretch}{1.2}
	\maketitle
	\footnotesize
	\captionof{table}{Summary of parameter values (if fixed) and bounds for use in the fitting procedure, including references. The fitted model is then implemented directly for the optimal vaccine allocation problem. To avoid overfitting, we limit the piecewise-constant domains for the contact reduction and keep the contact matrices constant (which were obtained from diaries).}
	\vspace{4pt}
	\label{tab:param_values}
	\begin{tabular}[b]{x{0.1\columnwidth}|x{0.35\columnwidth}|x{0.14\linewidth}|x{0.25\columnwidth}|x{0.12\columnwidth}}
		Param. & Description & Age-dependent & Value / Bound & Ref.  \\
		\hline
		$\gamma_e$ & Latent rate & No & $\gamma_e\in [0.15, 0.4]$ & ~\cite{LiPeiEtAl2020}\\
		$\gamma_i$ & Infectious rate & No & $\gamma_i\in [0.15, 0.4]$ &~\cite{LiPeiEtAl2020}  \\
		$\zeta_i$ & Relative susceptibility to infection & Yes & $\{u_i \geq u_j \ \text{if} \ i > j\}\ \wedge$ $\{u_i \in [0,1] \ \forall \ i\}$ & ~\cite{DaviesKlepacEtAl2020}\\
		$c_{i,j}$ & \# of contacts of ind. of group $i$ with one of group $j$ & Yes & Country-specific contact matrix &~\cite{MossongHensEtAl2008,PremCookEtAl2017}  \\
		$\eta_{\mathrm{s},i}^1$ & 1st-dose eff. in reducing susceptibility & Yes & 0.5 &~\cite{iso_ECDC:2021}  \\
		$\eta_{\mathrm{s},i}^2$ & 2nd-dose eff. in reducing susceptibility & Yes & 0.9&~\cite{iso_ECDC:2021}  \\
		$\eta_{\text{inf},i}^1$ & 1st-dose eff. in reducing infectiousness & Yes & 0.2&~\cite{iso_ECDC:2021}  \\
		$\eta_{\text{inf},i}^2$ & 2nd-dose eff. in reducing infectiousness & Yes & 0.2&~\cite{iso_ECDC:2021}  \\
		$\theta(t)$ & Contact reduction & No & $\theta(t)\in [0,1] \ \forall \ t$ & \\ 		
	\end{tabular}\vspace{10pt}
	\includegraphics[width=\linewidth]{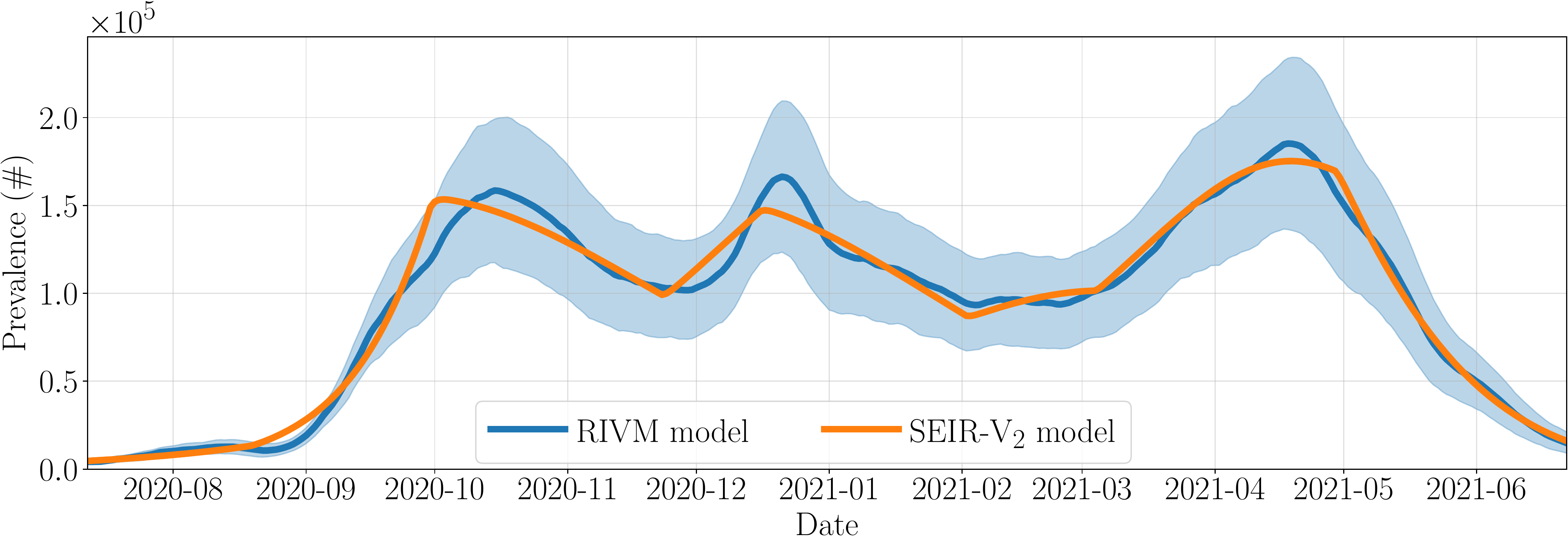}
	\captionof{figure}{Comparison of the overall prevalence of the high-fidelity stochastic RIVM model (with $95\%$ bounds) (blue), compared to the fit of our deterministic model (orange). We observe that our model captures the principal dynamics of the disease spread, notably capturing changes in behavior (due to government policy changes) and the effect of the vaccinations (the last 6 months).}
	\label{fig:model_fit}
	\vspace{-20pt}
\end{figure}

The bounds and/or values of parameters in Eq.~\eqref{eq:dynamics} are provided in Table~\ref{tab:param_values}. The bounds of the parameters are based on previous literature studies, as referenced in Table~\ref{tab:param_values}. We fix the contact matrix and the vaccine efficacy rates to avoid overfitting. The contact matrix is obtained from the European POLYMOD survey, which used population-based contact diaries to measure contacts in Western-European countries~\cite{MossongHensEtAl2008}, which has since been projected to other countries~\cite{PremCookEtAl2017}. Whilst it is somewhat limiting to fix the vaccine efficacy rates, given that different groups received doses at different values of $\theta(t)$, i.e., under different rules on contact reduction, fitting the vaccine efficacy rates can result in overfitting. Finally, the contact reduction parameter $\theta(t)$ describes the fraction of normal contacts that occur. It is considered piecewise constant and we assume they are correlated with the stringency of governmental restrictions. Specifically, we fix the change-points at dates of a large change in the severity of the measures imposed by the government based on a composite index that quantifies government restrictions across countries~\cite{HaleAngristEtAl2021}.
Fig.~\ref{fig:model_fit} shows that the proposed model can clearly capture the dynamic behavior of the pandemic.

	\section{Optimizing Vaccine Allocation}\label{sec:optimization_results}
\begin{figure*}[t]
	\centering
	\begin{minipage}{0.49\linewidth}
		\centering
		\includegraphics[width=\linewidth]{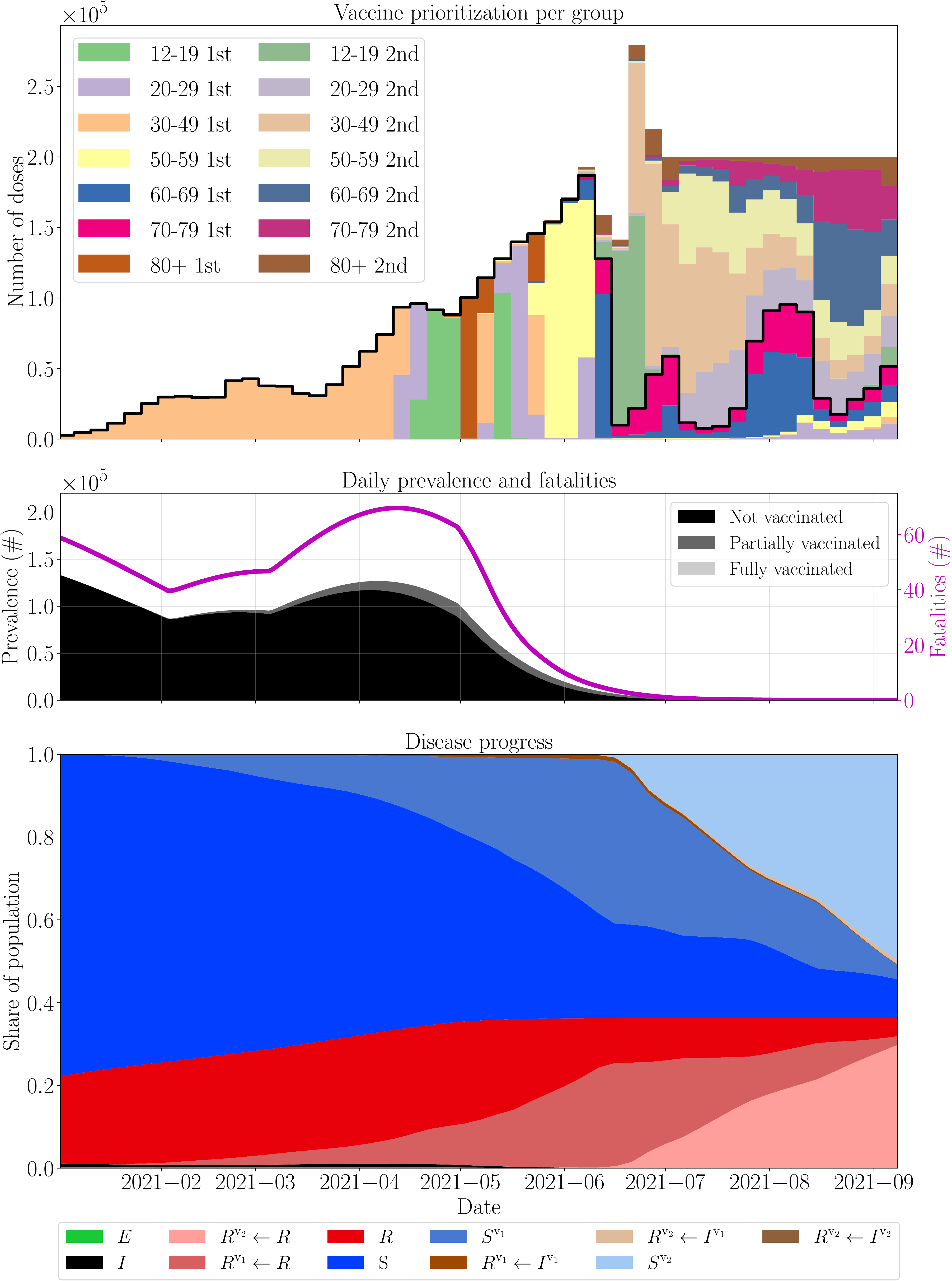}
		\captionof{figure}{\textbf{Minimizing infections:} Vaccine allocation (top, black solid line separating the shots), evolution of the number of people infected (middle), and overall disease dynamics evolution (bottom) when optimizing for minimizing total infections. The top figure shows the optimal strategy results in delaying second doses and prioritizing groups that have either (both) more overall contacts or (and) higher relative susceptibility. A moderate number of partially vaccinated people get infected.}
	\label{fig:cases}
	\end{minipage}%
	\hfill
	\begin{minipage}{0.49\linewidth}
		\centering
		\includegraphics[width=\linewidth]{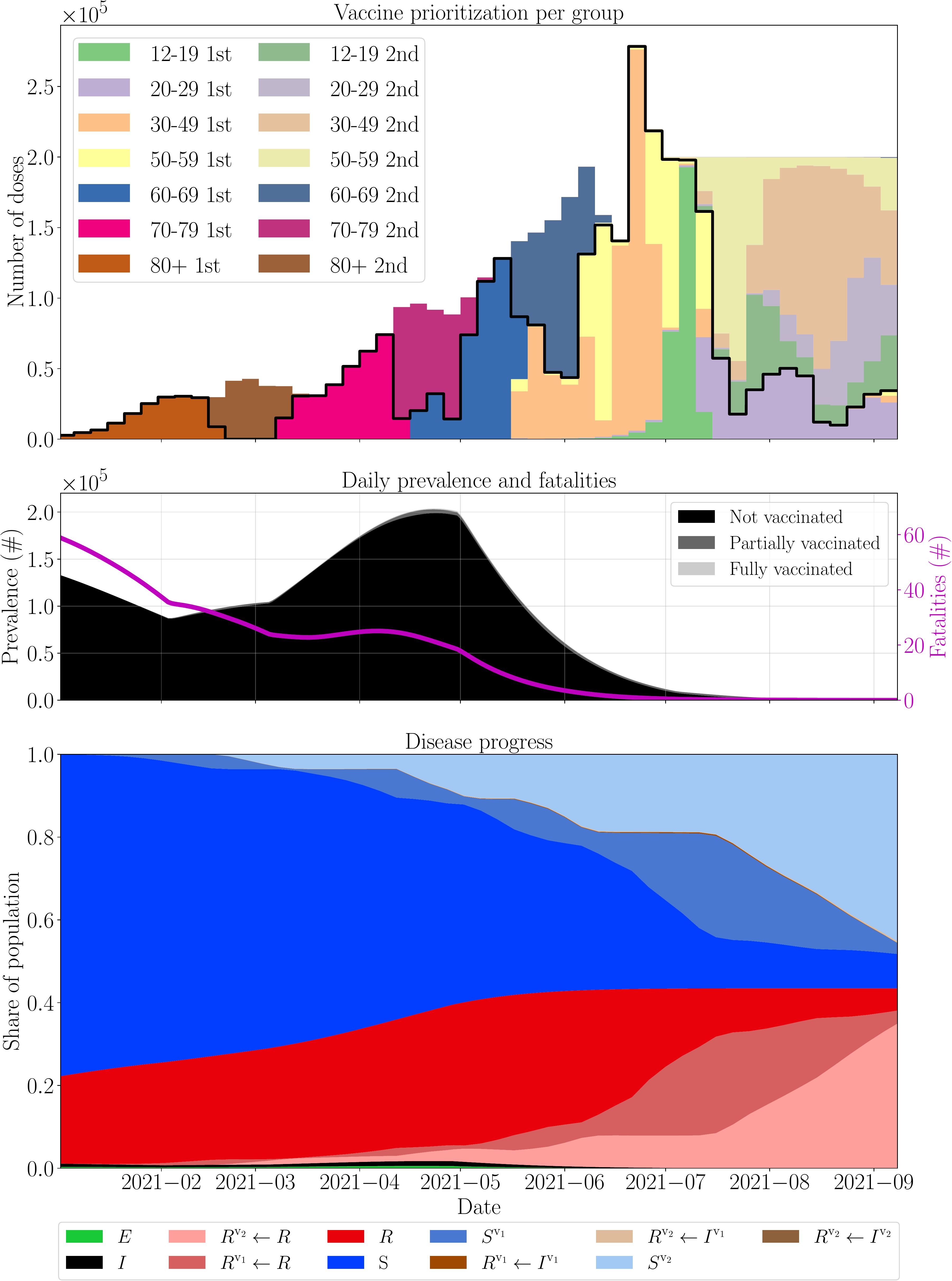}
		\captionof{figure}{\textbf{Minimizing fatalities:} Vaccine allocation (top, black solid line separating the shots), evolution of the number of people infected (middle), and overall disease dynamics evolution (bottom) when optimizing for minimizing total fatalities. The top figure shows the optimal strategy results in fully vaccinating groups that have high mortality rates, before proceeding with delaying second doses for younger age-groups, prioritizing suppressing the spread.
		Very few partially vaccinated people get infected.}
		\label{fig:deaths}
	\end{minipage}
\vspace{-.5cm}
\end{figure*}

The optimization problem~\eqref{eq:ocp} described in Section~\ref{subsec:ocp} is highly nonlinear.
We discretize Eq.~\eqref{eq:ocp} in time with Euler forward, parse the resulting static optimization problem with CasADi~\cite{AnderssonGillisEtAl2019} and solve it with IPOPT~\cite{WaechterBiegler2006}. Thereby, in an attempt to benchmark the quality of the solution found, we use multiple initializations. We provide an open-source implementation of our approach\footnote{\scriptsize\url{https://gitlab.tue.nl/20200365/covid-vaccine-allocation}}.

\subsection{Optimizing the Vaccine Rollout for Different Objectives}

\begin{table}[t]
	\centering
	\setlength\tabcolsep{0pt}
	\renewcommand{\arraystretch}{1.5}
	\maketitle
	\caption{Relative difference between the outcome stemming from a particular strategy (rows) and the achievable minimum obtained optimizing for that particular objective (columns).
		The results demonstrate that the local optima found for each strategy perform best on the metric for which they optimize, dominating the other objectives. Additionally, it demonstrates that optimizing for minimizing cumulative hospitalizations and fatalities is very similar, and mostly in line with the strategy of The Netherlands. All strategies outperform the baseline of proportional allocation and the allocation strategy in The Netherlands.}
	\label{tab:outcomes}
	\begin{tabular}{p{0.25\columnwidth}|*{4}{x{0.185\columnwidth}}}
		\diagbox[width=6em, height=1.8em]{$J$}{\raisebox{0.5\height}{Outcome \hspace{-16pt}}} & $\Delta$hosp\textsubscript{tot} &$\Delta$fatalities\textsubscript{tot} & $\Delta$infections\textsubscript{tot} & $\Delta$hosp\textsubscript{max} \\\hline
		min hosp\textsubscript{tot}& \cellcolor{gray} 0\%& +25 \%  & +14\%  & +5\% \\
		min fatalities\textsubscript{tot}& +6\% &   \cellcolor{gray} 0\%& +42\% & +13\% \\
		min infections\textsubscript{tot}& +11\% & +84\% & \cellcolor{gray}0\% & +26\% \\
		min hosp\textsubscript{max}& +2\% & +25\% & +15\% &  \cellcolor{gray}0\%\\
		Baseline\footnotemark& +19\% & +91\% & +17\% & +37\% \\
		The Netherlands\footnotemark& +8\% & +34\% & +28\% & +17\% \\
	\end{tabular}
	\vspace{-10pt}
\end{table}
\footnotetext[2]{Considering a proportional allocation with a dosing interval of 3 weeks.}
\footnotetext[3]{Compared until the end of the data fitting period, June 30th.}

As was done in The Netherlands, the vaccination campaign starts in early January 2021. For the first 6 months, we fix the maximum vaccine supply to the number of vaccine doses administered in The Netherlands~\cite{RIVM:2021b}, and consider an additional 2-month period in which a constant supply of vaccines was available. During this summer period, we consider less restrictive measures, as was implemented by The Netherlands. 
Table~\ref{tab:outcomes} reports the relative outcomes on different metrics for each optimized strategy. We also compare these results with the vaccine rollout in The Netherlands and a proportional allocation baseline in which second doses are given as per the manufacturers' recommendations (3 weeks).
Fig.~\ref{fig:cases} and \ref{fig:deaths} show strategies and pandemic evolution when minimizing total infections and fatalities, respectively.

\subsection{Discussion}
As can be seen from Fig.~\ref{fig:cases} and~\ref{fig:deaths}, different objectives (minimize infections and fatalities, respectively) can lead to significantly different allocation strategies. Fig.~\ref{fig:cases} shows that minimizing for total infections results in prioritizing partially vaccinating the large majority of the population before administering second doses. Within this prioritization, groups with high contact rates (young working age and student groups) and those with higher susceptibility (80+ year olds) are prioritized.
As we assume that a partially vaccinated individual sees a reduction in susceptibility that is more than half the reduction in susceptibility from full vaccination (see Table~\ref{tab:param_values}), prioritizing partial vaccinations makes intuitive sense. However, due to the prioritization of partially vaccinating a large proportion of the population, a moderate amount of partially vaccinated individuals still get infected, as can be seen in the bottom two subfigures of Fig.~\ref{fig:cases}.
In contrast to this strategy, minimizing the total number of fatalities results in prioritizing fully vaccinating the elderly populations (old-to-young), as shown in Fig.~\ref{fig:deaths}.
Once the 60+ categories have been vaccinated, priority is shifted to partially vaccinating the younger age groups, as their fatality rates are comparable (and very low). Compared to minimizing for the total number of infections, this strategy ensures that almost no partially vaccinated people are infected.
In both Fig.~\ref{fig:cases} and~\ref{fig:deaths}, we can appreciate the fact that after July the pandemic dies out, as herd immunity is reached. From this point onward, the vaccine allocation strategy has little effect on cumulative outcomes, resulting in strategies that combine various groups at the same time instance. This is also due to the fact that we are assuming immunity to be immediate and not to wane, and that vaccine hesitancy is implemented for the youngest category only, to model the unavailability of vaccines for children younger than 12 in the time-frame under consideration, whilst we assume that the whole population is willing to be vaccinated.

To characterize the results on a quantitative level, Table~\ref{tab:outcomes} compares the outcomes of the different optimized strategies on the cumulative fatalities and infections, and the cumulative and maximum hospital occupancy with the achievable minimum obtained when optimizing for that particular objective.
Thereby, we also include a baseline comparison in which the vaccine is distributed equally over the full population\footnotemark[2] and to the vaccination rollout that occurred in The Netherlands\footnotemark[3].
The baseline strategy performs significantly worse than the optimized strategies in each metric, highlighting the importance of optimizing the vaccine administration process.
In contrast, the approach chosen by The Netherlands strikes a trade-off between minimizing fatalities and infections, and is solely dominated by both the minimum-hospitalization strategies (cumulative and maximum).
Hereby, it should be noted that The Netherlands chose to prioritize vaccinating a proportion of essential healthcare workers during the early phase of the rollout, which we did not account for in our results.
Finally, Table~\ref{tab:outcomes} shows that each optimized strategy performs best on the metric for which they optimize, hence dominating the other strategies. Whilst not providing global optimality guarantees, this result clearly indicates that the found local optima are promising.

	\section{Conclusion}\label{sec:conclusion}
This paper investigated the vaccine allocation problem when fighting a pandemic outbreak that evolves dynamically within an age-stratified population.
To this end, we devised a model and an optimization framework to compute and characterize the optimal allocation strategies that minimize predefined societal objectives.
Specifically, we instantiated a finite-horizon optimal control problem based on a computationally-tractable epidemiological model, which we designed to account for partially- and fully-vaccinated people belonging to different age-groups, and which we fitted to realistically represent the COVID-19 prevalence data in The Netherlands.
Since compartmental epidemiological models are inherently non-convex, we solved the optimal control problem using nonlinear programming to a local optimum.
Our framework revealed that different societal objectives can lead to significantly different allocation strategies and total outcomes: The prioritization strategy found for minimizing the total number of infections demonstrated the necessity of explicitly modeling multi-dose vaccines in our SEIR-V\textsubscript{2} model, as it entails vaccinating the whole population with one dose first. In contrast, we showed that to minimize fatalities, the oldest part of the population must be fully vaccinated first.
Finally, we also observed that the found local optimum for each optimized strategy dominates the other optimized strategies and baselines on the metric for which it optimizes. 
Our framework can be used by policymakers to optimize, evaluate and compare different allocation strategies. Nonetheless, policymakers should be aware of its limitations: First, with this paper, we did not aim to address complex social and ethical considerations that can be socially disruptive, but rather focused on quantifiable scientific results.
Second, our conclusions are subject to a number of modeling assumptions that must be weighed in when interpreting the results.

Our work opens the field for the following extensions:
First, we plan to address the computational complexity of the large-scale nonlinear optimization problem using sequential convex programming.
Second, it would be useful to account for the uncertainty in the future evolution of the pandemic.
Finally, we want to study the interplay between policy measures, vaccinations, and people's behavior.
	
	\addtolength{\textheight}{-6.3cm}
	\section*{Acknowledgments}
	We thank Dr.\ M.\ Steinbuch and the M\'axima Medisch Centrum (MMC) for their support, and Ir.\ J.\ van den Hurk for assisting in validating the model and the quality of the local optima. We are grateful to Dr.\ I.\ New and T.\ Herrmann for proofreading this paper.

	 
	\bibliographystyle{IEEEtran}
	\bibliography{../../../Bibliography/main, ../../../Bibliography/SML_papers}
	
	\ifextendedversion
	\begin{appendices}
		\input{chapters/appendix}
	\end{appendices}
	\fi

\end{document}